\documentclass[aps,pra, twocolumn,nofootinbib]{revtex4-1}
\pdfoutput=1
\usepackage[utf8]{inputenc}
\usepackage{natbib}
\usepackage{amsmath}
\usepackage{amssymb}
\usepackage{braket}
\usepackage{bbm}

\newcommand\D[1]{\mathrm{D}\left(#1\right)}
\newcommand\DM[1]{\mathrm{D}_M\left(#1\right)}
\newcommand\DEm[1]{\mathrm{D}_\mathcal{M}\left(#1\right)}
\newcommand\deff[1]{{d}^{\mathrm{eff}}\left(#1\right)}
\newcommand\tr[1]{\mathrm{tr}\left(#1\right)}

\newcommand\trb[1]{\mathrm{tr}_B\left(#1\right)}

\newtheorem{theorem}{Theorem}

\begin{document}

\title{Eigenstate Thermalisation on Average}

\author{Joe Dunlop} \email{joe.h.dunlop@gmail.com}
\affiliation{H.H. Wills Physics Laboratory, University of Bristol, Tyndall Avenue, Bristol, BS8 1TL, U.K.}

\author{Oliver Cohen}\email{oc16271@alumni.bristol.ac.uk}
\affiliation{H.H. Wills Physics Laboratory, University of Bristol, Tyndall Avenue, Bristol, BS8 1TL, U.K.}

\author {Anthony J. Short} \email{tony.short@bristol.ac.uk}
\affiliation{H.H. Wills Physics Laboratory, University of Bristol, Tyndall Avenue, Bristol, BS8 1TL, U.K.}

\begin{abstract}

We consider conditions under which an isolated quantum system approaches a microcanonical equilibrium state. A key component is the eigenstate thermalisation hypothesis, which proposes that all energy eigenstates appear thermal. We introduce a weaker version of this requirement, applying only to the average distinguishability of eigenstates from the thermal state, and investigate its necessity and sufficiency for thermalisation.

\end{abstract}
\date{\today}
\maketitle

\section{Introduction}

If a confined physical system of many interacting bodies is displaced from equilibrium, it will typically relax to a thermal state that's well approximated by a statistical ensemble \cite{landau}. This empirical fact is the basis for equilibrium statistical mechanics, but quantum mechanical explanations for thermalisation remain an active topic of research \cite{Reimann_2008,Linden_2009,Short_2011,Deutsch_2018,biroli_2010,reimann_2018,ikeda_13,Iyoda_17,Riera_2012,muller_2013,Reimann_2010,dalessio_2016,Gogolin_2016,yoshizawa_18,Popescu_2006,tasaki_16,mori_2017}.

A first-principles treatment should recover thermalisation behaviour for typical pure states of an isolated quantum system, preferably under minimal assumptions. Isolated systems can be modelled exactly, require no additional assumptions about external interactions, and are arguably a general case: open systems can be treated as subsystems of a larger isolated system. An initial state with well-defined energy is expected to become indistinguishable from the relevant microcanonical state. How and why this should happen is not obvious: a pure state remains pure throughout its unitary evolution, and is in principle possible to discriminate from the mixed microcanonical state at all times.

However, the ideal measurement for distinguishing a given state from a thermal mixture is often practically unfeasible \cite{Reimann_2008}. Known conditions on the initial state and Hamiltonian guarantee equilibration with respect to \textit{limited} sets of measurements \cite{Reimann_2008,Linden_2009,Short_2011}. We demonstrate that a broad class of states which equilibrate - namely those which overlap significantly with a large number of energy levels - must also thermalise, provided that energy eigenstates are \textit{on average} difficult to distinguish from the microcanonical state. This is a significantly weaker requirement than the usual sense of eigenstate thermalisation \cite{Deutsch_2018}. We also consider to what extent eigenstate thermalisation is necessary for all such states to thermalise. Alternative weak versions of the eigenstate thermalisation hypothesis, focusing on expectation values of observables, have previously been considered \cite{biroli_2010,reimann_2018, ikeda_13,Iyoda_17}, and their necessity investigated using typicality arguments \cite{reimann_2018}.

A system is said to equilibrate if it approaches a fixed state (its long-term average), and stays close to it for almost all subsequent times. It thermalises if the time-averaged state is close to a microcanonical state, the uniform statistical mixture of all possible states satisfying the constraints of the system, particularly bounds on its total energy \cite{Linden_2009}. Our notion of closeness is provided by a distinguishability metric, expressing an ideal experimenter's ability to tell states apart using a fixed set of measurements.

 Restricted measurement capability is a fairly natural assumption, analogous to coarse-graining the phase space of a classical system into macrostates \cite{Reimann_2008,Short_2011}. For example, the allowed measurements might be those whose procedures and outcomes can be specified in a certain number of binary digits, most likely tiny in comparison to the degrees of freedom of a macroscopic system. A case of special interest is where measurements can be made only on a subsystem, with the remainder of the system treated as an unseen environment (heat bath). Under suitable assumptions of weak subsystem-bath coupling, the microcanonical state traced onto the subsystem is expected to take a Gibbs canonical form \cite{landau,Riera_2012,muller_2013,Reimann_2010}.

\section{Mathematical Preliminaries}

In general, we will consider measurements restricted to a fixed set $\mathcal{M}$ of positive operator valued measures (POVMs) $M$, each with a discrete set $\{M_r\}$ of outcome operators. The distinguishability of a pair of density operators $\rho$ and $\sigma$ with respect to the measurement set $\mathcal{M}$ is defined as

\begin{equation}
    \DEm{\rho,\sigma}:= \frac{1}{2}\max_{M\in\mathcal{M}}\sum_{r=1}^{N(M)}\left|\tr{M_r\left(\rho-\sigma\right)}\right|,
\end{equation}
and takes a value from 0 to 1. An experimenter with either state $\rho$ or $\sigma$ and able to make measurements from $\mathcal{M}$ can guess which state they were given with probability at most $\frac{1}{2}\left(1+\DEm{\rho,\sigma}\right)$. In the case that $\mathcal{M}$ corresponds to the set of all measurements on a subsystem S - that is, every POVM whose outcomes are of the product form $M_R^S\otimes\mathbb{I}^B$ - distinguishability is equivalent to trace distance on the subsystem,

\begin{equation}
    \D{\rho^S,\sigma^S}:= \frac{1}{2}\tr{\left|\rho^S-\sigma^S\right|},
\end{equation}

where $\rho^S=\trb{\rho}$ and $\sigma^S=\trb{\sigma}$  are reduced density operators.

We will consider systems with a discrete Hamiltonian $H = \sum_{n}E_n \rho_n$, where $\rho_n = \ket{n}\bra{n}$ denote the energy eigenstates, and for simplicity assume that there are no energy degeneracies\footnote{ In physical contexts, exact energy degeneracies may not pose a significant problem because they will typically be lifted by a slight random perturbation of the Hamiltonian.}. Modified versions of our theorems which apply to degenerate systems are presented in Appendix D.
In quantum mechanics, the microcanonical ensemble refers to a narrow \textit{band} of energies
. Given a band $\textrm{span}\{\rho_n : E\leq E_n \leq E+\Delta\}$, for which we will index the energy eigenstates $\rho_1,...,\rho_d$, the microcanonical state $\Omega$ is given by the uniform mixture:
\begin{equation}
    \Omega = \frac{1}{d}\sum_{n=1}^{d}\rho_n.
\end{equation}
Given any initial state $\rho(0)$ of the system (which may be pure or mixed), its infinite time average is given by the dephased diagonal operator
\begin{align}
   \omega &= \langle \rho(t) \rangle_t \nonumber \\
   &= \lim_{T \rightarrow \infty} \frac{1}{T} \int_0^{T} dt \,  e^{-\frac{i}{\hbar} H t} \rho(0)  e^{+\frac{i}{\hbar} H t} \nonumber \\ 
   &= \sum_{n} p_n \rho_n.
\end{align}
where $p_n = \langle n |\rho(0) |n \rangle$. 

The central objective of this paper is to investigate conditions on the initial state, Hamiltonian and measuring capability under which the time-averaged state is practically indistinguishable from the band's microcanonical state:
\begin{equation}
    \DEm{\omega,\Omega}\ll 1.
\end{equation}

\section{Equilibration and the Eigenstate Thermalisation Hypothesis}

It's known that under relatively weak assumptions, states with high \textit{effective dimension} equilibrate \cite{Linden_2009,Short_2011}. Effective dimension provides a weighted measure of the number of energy levels a state overlaps with. An invariant under unitary time evolution, the effective dimension of a state $\rho(t)$ with time-average $\omega = \sum p_n \rho_n$ is given by
\begin{equation}\label{deff}
    \deff{\omega}=\frac{1}{\sum_n p_n^2},
\end{equation}
provided the energy spectrum is nondegenerate.

For many relevant Hamiltonians, the density of energy levels in the bulk of the spectrum scales exponentially with the size of the system, meaning that the dimension $d$ of even a narrow energy band becomes extremely large ($\sim10^{10^{23}}$) for macroscopic systems \cite{Reimann_2010}. In a band with high dimension, the vast majority of pure states have $d^{\textrm{eff}}\geq\frac{d}{4}$, so any behaviour that holds for states whose effective dimension is an order-1 fraction of the band dimension can be said to hold for typical pure states \cite{Linden_2009}. Moreover, it is argued in \cite{Reimann_2010} that states which populate relatively few energy levels cannot be prepared under realistic experimental conditions. We will refer to states with $d^{\textrm{eff}}\geq\frac{d}{4}$ as having ``high effective dimension''.

Initial states entirely confined to a narrow energy band are clearly an idealisation: this can be accounted for by choosing a band such that for some suitably small $\delta$, the state's total overlap $\sum_{\textrm{band}} p_n \geq 1-\delta$, thus excluding the energy distribution's tails. Our results still hold to a good approximation after such an adjustment is made, as demonstrated in Appendix D.

If the Hamiltonian has nondegenerate energy gaps, the time-averaged distinguishability of a state $\rho(t)$ from its long-term average $\omega$ is bounded by
\begin{equation}
    \langle \DEm{\rho(t),\omega}\rangle _t \leq \frac{N_\mathcal{M}}{4\sqrt{\deff{\omega}}},
\end{equation}
where $N_\mathcal{M}$ is the total number of possible outcomes\footnote{If $\mathcal{M}$ represents measurements on a subsystem, $N_{\mathcal{M}}$ can be substituted for the subsystem dimension $d_S$ in the above result \cite{Linden_2009}.} across all measurements in $\mathcal{M}$ \cite{Short_2011}. As a result, any state with effective dimension much larger than $({N_\mathcal{M}})^2$ equilibrates with respect to $\mathcal{M}$. Provided the number of resolvable measurement outcomes does not scale with system size as fast as the density of energy levels, equilibration is more or less assured for realistic initial states of a large system.


There is no such clear cut reason that the equilibrium state $\omega$ should appear microcanonical, but a possible explanation lies in the eigenstate thermalisation hypothesis (ETH) \cite{Deutsch_2018,dalessio_2016}. Stated in terms of distinguishability, the standard ETH postulates that in complex interacting systems, all of the energy eigenstates in a given energy band are near-indistinguishable from the band's microcanonical state $\Omega$:
\begin{equation}\label{eq:ETH}
    \forall \rho_n:\;E\leq E_n \leq E+\Delta,\hspace{5mm} \DEm{\rho_n,\Omega} \ll 1.
\end{equation}
A pure superposition state in the band might initially be easily distinguished from $\Omega$ due to special phase relationships, but the time-averaged state will be no more distinguishable from the microcanonical state than the least-thermalised energy eigenstate:
\begin{equation}\label{eq:ETH_2}
\begin{split}
    \DEm{\omega,\Omega} &= \DEm{\sum_n p_n \rho_n,\Omega}\\
    &\leq\sum_n p_n \DEm{\rho_n,\Omega}\\
    &\leq \max_n \DEm{\rho_n,\Omega}\\
    &\ll 1,
\end{split}
\end{equation}
where we have used the assumption \eqref{eq:ETH} in the final step.
Conversely, if we demand that every equilibrating state thermalises, eigenstate thermalisation must hold because energy eigenstates are trivially equilibrated (since they are time-independent). Thermalisation of all energy eigenstates is necessary and sufficient condition for all time-averaged states in an energy band to appear thermal \cite{Gogolin_2016}.

The eigenstate thermalisation hypothesis emerged from the study of chaotic semiclassical systems \cite{deutsch_1991,Srednicki_1994}, and is usually framed as a condition on the diagonal energy-basis matrix elements of `realistic' observables. A number of numerical and analytical studies have demonstrated eigenstate thermalisation, particularly for few-body observables on locally interacting lattices, and quantum many-body systems which are chaotic in the semiclassical limit \cite{dalessio_2016,yoshizawa_18}. 
The ETH is also supported by the principle of \textit{canonical typicality}: the vast majority of randomly selected pure states appear thermal with respect to measurements on a subsystem of a sufficiently large quantum system \cite{goldstein_05,Popescu_2006,tasaki_16}. There has been limited consideration of eigenstate thermalisation in the context of distinguishability metrics, which provides the framework to account for more general measurement capabilities \cite{Gogolin_2016}.

There has recently been some discussion of weaker conditions which still lead to thermalisation of a majority of equilibrating states \cite{biroli_2010,reimann_2018,ikeda_13,Iyoda_17}, and questions of whether the ETH is \textit{needed} to explain observed thermalisation behaviour \cite{mori_2017,reimann_2018}. This has in part been motivated by the identification of \textit{quantum many-body scars} in condensed matter systems, referring to collections of ETH-violating energy eigenstates (normally a small fraction of the system dimension) \cite{Serbyn2021,Turner2018}. The presence of a scar can lead to large scale revivals for pure states which have significant support in the non-thermal eigenstates, although most equilibrating states will still thermalise \cite{alhambra_20}. We introduce a weakening of the ETH based on the intuitive notion of average distinguishability between the individual eigenstates and the microcanonical state, which we call \textit{eigenstate thermalisation on average}. We then explore quantitatively  the extent to which this is necessary and sufficient for the thermalisation of all states with high effective dimension.

\section{Eigenstate thermalisation on average}

Given the version of the Eigenstate Thermalisation Hypothesis presented above \eqref{eq:ETH}, time-averaged states need only be as close to the microcanonical state as the least-thermal energy eigenstate. 
However, for macroscopic systems, the density of energy eigenstates is enormous. In such situations, it's desirable to base results on properties of the bulk of eigenstates, rather than of a one-in-$2^{10^{23}}$ outlier.

A state cannot meaningfully thermalise if it does not also equilibrate: it doesn't matter whether the time-averaged state is close to the relevant thermal state if the instantaneous state doesn't approach its time average. Because of this, it makes most sense to examine conditions for thermalisation in systems which are known to equilibrate.

The most important factor in equilibration of closed systems is the state's effective dimension. High effective dimension expresses that a state has overlap with a large number of energy levels, and conversely that the overlap with any single energy eigenstate is relatively small. If only a few eigenstates are easy to distinguish from the band mixture, these will have little effect on the superposition \cite{biroli_2010,reimann_2018}. This is the essence of our first result, which demonstrates that all states with high enough effective dimension must thermalise if the average distinguishability of energy eigenstates from the band mixture is suitably small. We denote the mean and root mean square distinguishabilities by

\begin{equation}\label{Dmean}
    \mathrm{D}_{\mathcal{M}}^{\mathrm{mean}} = \frac{1}{d}\sum_{n=1}^{d} \DEm{\rho_n,\Omega}
\end{equation}
and
\begin{equation}\label{DRMS}
    \mathrm{D}_{\mathcal{M}}^{\mathrm{RMS}} = \sqrt{\frac{1}{d}\sum_{n=1}^{d} (\DEm{\rho_n,\Omega})^2}
\end{equation}
respectively.
\vspace{1em}
\begin{theorem}

Let $\rho_1,...,\rho_d$ be energy eigenstates spanning a narrow energy band, and let $\Omega = \frac{1}{d}\sum_n\rho_n$ denote the band's microcanonical state. The distinguishability of any time-averaged state $\omega\in\textrm{span}\{\rho_1,...,\rho_d\}$ from the microcanonical state is bounded by
\begin{equation}
    \begin{split}
    \DEm{\omega,\Omega} &\leq \mathrm{D}_{\mathcal{M}}^{\mathrm{RMS}}\sqrt{\frac{d}{\deff{\omega}}-1}\\
    &\leq \sqrt{\mathrm{D}_{\mathcal{M}}^{\mathrm{mean}}\left(\frac{d}{\deff{\omega}}-1\right)},
    \end{split}
\end{equation}
where $\mathrm{D}_{\mathcal{M}}^{\mathrm{mean}}$ and $\mathrm{D}_{\mathcal{M}}^{\mathrm{mean}}$ are respectively the mean and root mean square distinguishability of energy eigenstates from the microcanonical state with respect to the measurement set $\mathcal{M}$.

\end{theorem} 
\vspace{1em}

A proof is provided in appendix A. As a consequence of theorem 1, in any energy band where eigenstate thermalisation holds on average, in that $\mathrm{D}^\mathrm{RMS}_\mathcal{M} \ll1$ or $\sqrt{\mathrm{D}^\mathrm{mean}_\mathcal{M}} \ll 1$, any initial state with high effective dimension must be such that its time average is near indistinguishable from the microcanonical state $\Omega$.
If the state's effective dimension is at least $\frac{d}{4}$, the distinguishability of its long-term average from $\Omega$ is no greater than $\sqrt{3\, \mathrm{D}^\mathrm{mean}_\mathcal{M}}$.

\section{How far is eigenstate thermalisation necessary?}

We have seen that eigenstate thermalisation on average ensures that all states with high effective dimension become indistinguishable from the microcanonical state. It is natural to ask whether the same condition is \textit{necessary} for those states to thermalise. Our second result bounds the mean eigenstate thermalisation in terms of the minimum degree of thermalisation for high effective dimension states:

\vspace{1em}
\begin{theorem}
Suppose that, in an energy band with dimension $d$, every time-averaged state $\omega$ with $\deff{\omega}\geq\frac{d}{4}$ is such that $\DEm{\omega,\Omega}\leq\varepsilon$ for some $\varepsilon>0$.

\renewcommand{\labelenumi}{\roman{enumi}}
\begin{enumerate}
    \item If the number of outcomes across all available measurements, $N_{\mathcal{M}}=\sum_{M\in\mathcal{M}}N(M)$ is finite, then the mean distinguishability of energy eigenstates from the band's microcanonical state,
\begin{equation}\label{finitebound}
    \frac{1}{d} \sum_{n=1}^{d}\DEm{\rho_n,\Omega}\leq N_{\mathcal{M}}\,\varepsilon.
\end{equation}
    \item{Furthermore, if $\mathcal{M}$ represents the set of all POVMs on a subsystem of dimension $d_\textrm{S}$, the mean distinguishability is bounded by
\begin{equation}\label{subsysbound}
    \frac{1}{d} \sum_{n=1}^{d}\DEm{\rho_n,\Omega}\leq {d_\textrm{S}}^\frac{5}{2}\varepsilon.
\end{equation}}
\end{enumerate}
\end{theorem}

See appendix B for proof. In effect, eigenstate thermalisation necessarily holds on average in an energy band if every state with high effective dimension thermalises strongly or if relatively few measurement outcomes are possible. The result is particularly powerful when applied to a very small subsystem, for example a qubit coupled to an unseen environment: in such cases, eigenstate thermalisation on average is necessary for the subsystem to approach a thermal state for all high effective dimension states of the full isolated system.

Because of the bounds' dependence on measuring power (as expressed through number of outcomes or subsystem dimension), it's not possible to say with generality that eigenstate thermalisation is strictly necessary for all high effective dimension states to thermalise. The following example demonstrates that when a very large number of measurements are available, a system might display strong thermalisation even when every energy eigenstate is easily distinguished from the microcanonical state. The intuition is that if the number of outcomes is comparable to the dimension of the energy band, then measurements might be `fine-tuned' to identifying individual energy eigenstates, but still fail to distinguish mixtures of a large number of eigenstates from the microcanonical state.
\vspace{1em}

\textbf{Example}
Consider an energy band spanned by eigenstates $\rho_1,...,\rho_d$ and suppose that an observer is able to make measurements from the set $\mathcal{M}=\{M_1,...,M_d\}$, where each $M_n$ has binary outcomes $M_n^+ = \rho_n$ and $M_n^- = \mathbb{I}-\rho_n$. For each energy eigenstate, 
\begin{equation}
\begin{split}
    \DEm{\rho_n,\Omega} & = \frac{1}{2}\max_{m}\Big[\big|\tr{\rho_m(\rho_n-\Omega)}\big|\\
    & \hspace{13mm} +\big|\tr{(\mathbb{I}-\rho_m)(\rho_n-\Omega)}\big|\Big]\\
    & =\max_{m}\big|\tr{\rho_m(\rho_n-\Omega)}\big|\\
    & = 1-\frac{1}{d},
\end{split}
\end{equation}
and as a result, the mean distinguishability of energy eigenstates from the microcanonical state is
\begin{equation}
    \frac{1}{d}\sum_{n=1}^{d}\DEm{\rho_n,\Omega}=1-\frac{1}{d}.
\end{equation}
In the limit of large $d$, each $\rho_n$ is almost perfectly identifiable by the appropriate measurement; there is no eigenstate thermalisation with respect to $\mathcal{M}$. Now, consider the generic time-averaged state $\omega = \sum_{n}p_n \rho_n$. Its distinguishability from the microcanonical state is given by:
\begin{equation}
    \begin{split}
    \DEm{\omega,\Omega} & = \frac{1}{2}\max_{m}\Big[\big|\tr{\rho_m(\omega-\Omega)}\big|\\
    & \hspace{13mm} +\big|\tr{(\mathbb{I}-\rho_m)(\omega-\Omega)}\big|\Big]\\
    & = \max_{m}\big|\tr{\rho_m(\omega-\Omega)}\big|\\
    & = \max_{m}\left|\tr{\rho_m\left(\sum_n\left(p_n  - \frac{1}{d}\right)\rho_n\right)}\right|\\
    & = \max_{m}\left|p_m - \frac{1}{d}\right|\\
    & \leq \sqrt{\sum_{n=1}^d \left(p_n - \frac{1}{d}\right)^2}\\
    & = \sqrt{\sum_{n=1}^{d} \left(p_n^2- 2\,\frac{ p_n}{d} + \frac{1}{d^2}\right)}\\
    & = \sqrt{\frac{1}{\deff{\omega}} - \frac{1}{d}}\\
    & \leq \frac{1}{\sqrt{\deff{\omega}}}.
    \end{split}
\end{equation}
In the penultimate line, we used the formula for effective dimension (\ref{deff}) and the fact that $\sum_{n=1}^{d}p_n = 1$.

Any time-averaged state with $d^{\textrm{eff}} \gg 1$ will be effectively indistinguishable from the microcanonical state by any of the measurements in $\mathcal{M}$. Moreover, any state of the system appears perfectly equilibrated at all times, since the outcome probabilities for all available measurements depend solely on the absolute values of the state's energy basis coefficients, which are time invariant.

While the measurement set considered here represents a logical extreme with little bearing on what's possible in the laboratory, it nonetheless provides a clear counterexample to any sweeping claim that eigenstate thermalisation, on average or otherwise, is always strictly necessary if all high effective dimension states are to thermalise.
\section{Conclusion}
It is broadly recognised that eigenstate thermalisation is a key condition for the thermalisation of isolated quantum systems, but it remains unclear to what degree the hypothesis holds in physical contexts. Much of the recent discussion favours a strong ETH, placing conditions on every energy eigenstate in order to justify thermalisation of all equilibrating states.

We find that eigenstate thermalisation on average is still sufficient for the great majority of pure initial states to thermalise, while being tolerant of a small number of exceptional, `far-from-thermal' eigenstates in a system's spectrum, as is characteristic of many-body scars. The shared property of the initial states known to thermalise, high effective dimension, is already known to guarantee equilibration.

In addition, when all states with high effective dimension thermalise strongly in comparison to the number of possible measurement outcomes, eigenstate thermalisation necessarily holds on average.

However, when a very large number of outcomes are possible, there are cases where all high effective dimension states thermalise even when eigenstate thermalisation does not hold \textit{at all}. It is unclear whether such a situation can be expected to arise in practice.

An interesting aspect of our results is their reliance on the ratio of the initial state's effective dimension to the dimension of an arbitrarily chosen energy band, reflecting that the definition of a microcanonical state is ultimately an operational one. An interesting direction for future investigation might be to choose an appropriate band depending on the initial state.

\bibliography{refs}
\appendix
\section{Proof of Theorem 1}\label{ap:thrm1}
Let $\omega$ be a time-averaged state $\sum_{n=1}^{d} p_n \rho_n$. With respect to any measurement $M\in\mathcal{M}$, the distinguishability of $\omega$ from the microcanonical state $\Omega$ is given by
\begin{equation} \label{eq:thm1proof} 
    \begin{split}
    &\DM{\omega,\Omega} = \frac{1}{2}\sum_{r=1}^{N(M)}\left|\tr{M_r\left(\omega-\Omega\right)}\right|\\
    &= \frac{1}{2}\sum_r\left|\tr{M_r\left(\sum_n\left(p_n -\frac{1}{d}\right)\left(\rho_n-\Omega\right)\right)}\right|\\
    &= \frac{1}{2}\sum_r\left|\sum_n\left(p_n -\frac{1}{d}\right)\tr{M_r\left(\left(\rho_n-\Omega\right)\right)}\right|\\
    \end{split}
\end{equation}
\begin{equation}
    \begin{split}         
    &\leq \frac{1}{2}\sum_r\sum_n\left|p_n -\frac{1}{d}\right|\left|\tr{M_r\left(\left(\rho_n-\Omega\right)\right)}\right|\\
    &= \sum_n\left|p_n -\frac{1}{d}\right|\left(\frac{1}{2}\sum_r\left|\tr{M_r\left(\left(\rho_n-\Omega\right)\right)}\right|\right)\\
    &= \sum_n\left|p_n -\frac{1}{d}\right|\,\DM{\rho_n,\Omega}\\
    &\leq \sum_n\left|p_n -\frac{1}{d}\right|\,\DEm{\rho_n,\Omega}\\
    &\leq \sqrt{\sum_n\left(p_n -\frac{1}{d}\right)^2}\,\sqrt{\sum_n\left(\DEm{\rho_n,\Omega}\right)^2}\\
    &= \sqrt{\sum_{n=1}^d\left(p_n^2 -2\,\frac{p_n}{d}+\frac{1}{d^2}\right)}\,\sqrt{\sum_n\left(\DEm{\rho_n,\Omega}\right)^2}\\
    &= \sqrt{\frac{1}{\deff\omega}-\frac{1}{d}}\;\sqrt{\sum_n\left(\DEm{\rho_n,\Omega}\right)^2}.\\
    \end{split}
\end{equation}

Since the above holds for any and every measurement $M\in\mathcal{M}$, it follows that
\begin{equation}
    \DEm{\omega,\Omega} \leq \sqrt{\frac{1}{d}\sum_n\left(\DEm{\rho_n,\Omega}\right)^2}\;\sqrt{\frac{d}{\deff\omega}-1},
\end{equation}
which is tighter of the claimed bounds. Furthermore, since $0\leq \DEm{\rho_n,\Omega} \leq 1$ for each $n$, it holds that
\begin{equation}
    \DEm{\omega,\Omega} \leq \sqrt{\frac{1}{d}\sum_n\DEm{\rho_n,\Omega}}\;\sqrt{\frac{d}{\deff\omega}-1},
\end{equation}
which is the looser bound.
\begin{flushright}
$\square$
\end{flushright}

\section{Proof of Theorem 2}\label{ap:thrm2}
\textbf{Proof of Theorem 2(i)}

Consider a time-averaged state $\omega$ in the energy band $\textrm{span}\{\rho_1,...,\rho_d\}$, with the form $\omega = \frac{1}{k}\sum_{j=1}^{k}\rho_{n_j}$ for some integer $1\leq k\leq \frac{d}{3}$. Note that for this state $\deff{\omega}=k$. The distinguishability of this state from the band's uniform mixture $\Omega$ by a single measurement $M$ is given by:

\begin{equation}
\begin{split}
    \DM{\omega,\Omega} & =  \frac{1}{2}\sum_{r=1}^{N(M)}\left| \tr{M_r\left(\left(\frac{1}{k}\sum_{j=1}^{k}\rho_{n_j}\right)-\Omega\right)}\right|\\
    & = \frac{1}{2}\sum_{r=1}^{N(M)}\left| \tr{M_r\left(\frac{1}{k}\sum_{j=1}^{k}\left(\rho_{n_j}-\Omega\right)\right)}\right|\\
    & = \frac{1}{2k}\sum_{r=1}^{N(M)}\left| \sum_{j=1}^{k}\tr{M_r\left(\rho_{n_j}-\Omega\right)}\right|\\
    \end{split}
\end{equation}
\begin{equation}
    \begin{split}     
    & \geq \frac{1}{2k}\left| \sum_{j=1}^{k}\tr{M_1\left(\rho_{n_j}-\Omega\right)}\right|.
\end{split}
\end{equation}

Where we assume without loss of generality that 
\begin{equation} \sum_{n=1}^{d}\left|\tr{M_1\left(\rho_{n}-\Omega\right)}\right| \geq \frac{1}{N(M)}\sum_{r=1}^{N(M)} \sum_{n=1}^{d}\left|\tr{M_r\left(\rho_{n}-\Omega\right)}\right|.
\end{equation}
By the lemma (appendix C), it is possible to choose the subset of eigenstates $\{\rho_{n_1},...,\rho_{n_k}\}$ such that
\begin{equation}
\left| \sum_{j=1}^{k}\tr{M_1\left(\rho_{n_j}-\Omega\right)}\right| \geq \frac{k}{d}\sum_{n=1}^{d}\left|\tr{M_1\left(\rho_{n}-\Omega\right)}\right|,
\end{equation}
and doing so,

\begin{equation}
\begin{split}
    \DM{\omega,\Omega} &\geq  \frac{1}{2k}\cdot\frac{k}{d}\sum_{n=1}^{d}\left|\tr{M_1\left(\rho_{n}-\Omega\right)}\right|\\
    &\geq \frac{1}{N(M)}\cdot\frac{1}{d}\sum_{n=1}^{d}\frac{1}{2}\sum_{r=1}^{N(M)}\left|\tr{M_r\left(\rho_{n}-\Omega\right)}\right|\\
    &= \frac{1}{N(M)}\cdot\frac{1}{d}\sum_{n=1}^{d}\DM{\rho_n,\Omega}.
    \end{split}
\end{equation}

To reiterate, given a measurement $M$ and an integer $k\leq \frac{d}{3}$, it is possible to construct a state with effective dimension exactly $k$ such that the distinguishability of that state's time-average from the maximally mixed state, $\DM{\omega, \Omega} \geq \frac{1}{N(M)}\cdot\frac{1}{d}\sum_{n=1}^{d}\DM{\rho_n,\Omega}.$ We choose $k$ such that $k\geq \frac{d}{4}$.

By extension, given a \textit{set} of measurements $\mathcal{M}$, it is possible to choose such a state with

\begin{equation}
\begin{split}
    \DEm{\omega,\Omega} & = \max_{M\in\mathcal{M}} \DM{\omega,\Omega}\\
    &\geq \max_{M\in\mathcal{M}} \frac{1}{N(M)}\cdot\frac{1}{d}\sum_{n=1}^{d}\DM{\rho_n,\Omega}.
\end{split}
\end{equation}

Noting that any weighted average of a set is no greater than the maximal value,

\begin{equation}
\begin{split}
    \DEm{\omega,\Omega} 
    &\geq \sum_{M\in\mathcal{M}}\frac{N(M)}{N_{\mathcal{M}}}\cdot\frac{1}{N(M)}\cdot\frac{1}{d}\sum_{n=1}^{d}\DM{\rho_n,\Omega}\\
    &= \frac{1}{N_{\mathcal{M}}} \cdot\frac{1}{d}\sum_{n=1}^{d}\sum_{M\in\mathcal{M}}\DM{\rho_n,\Omega}\\
    &\geq \frac{1}{N_{\mathcal{M}}} \cdot\frac{1}{d}\sum_{n=1}^{d}\DEm{\rho_n,\Omega},
\end{split}
\end{equation}

which proves the contrapositive of the claim: if $\frac{1}{d}\sum_{n=1}^{d}\DEm{\rho_n,\Omega} > N_{\mathcal{M}}\,\varepsilon$, then it would be possible to construct a state with effective dimension at least $\frac{d}{4}$ such that $\DEm{\omega,\Omega}>\varepsilon$.
\begin{flushright}
$\square$
\end{flushright}

\textbf{Proof of Theorem 2(ii)}

This proof follows a similar structure to that of Theorem 2(i): we show that there must exist a time-averaged state of the form $\frac{1}{k}\sum_{j=1}^k\rho_{n_j}$, with effective dimension $\frac{d}{4}\leq k\leq\frac{d}{3}$, such that the trace distance $\D{\omega^S,\Omega^S}\geq (d_S)^{-\frac{5}{2}}\sum_{n=1}^d\D{\rho_n^S,\Omega^S}$ on the subsystem. We introduce a Hermitian orthonormal operator basis $\{e_1,...,e_{d_S^2}\}$ on the subsystem such that $\tr{e_i e_j} = \delta_{ij}$, and for Hermitian operators $A$, $\tr{A\,e_i} \in \mathbb{R}$. In the seventh line below, we use the lemma (Appendix C) to lower bound the Hilbert-Schmidt projection of $(\omega^S-\Omega^S)$ onto the $e_1$ basis operator, where we assume without loss of generality that for $i>1$, $\sum_{n=1}^d\left|\tr{\left(\rho_{n}^S-\Omega^S\right)\,e_1}\right|\geq\sum_{n=1}^d\left|\tr{\left(\rho_{n}^S-\Omega^S\right)\,e_i}\right|$.

\begin{equation}
\begin{split}
    \D{\omega^S,\Omega^S}&=\frac{1}{2}\mathrm{tr}\left|\omega^S-\Omega^S\right|\\
    &=\frac{1}{2}\tr{\sqrt{\left(\omega^S-\Omega^S\right)^2}}\\
    &\geq\frac{1}{2}\sqrt{\tr{\left(\omega^S-\Omega^S\right)^2}}\\
    &=\frac{1}{2}\sqrt{\sum_{i=1}^{d_S^2}\left[\tr{\left(\omega^S-\Omega^S\right)\,e_i}\right]^2}\\
    &\geq\frac{1}{2}\,\left|\tr{(\omega^S-\Omega^S)\,e_1}\right|\\
    &=\frac{1}{2k}\,\left|\sum_{j=1}^k\tr{\left(\rho_{n_j}^S-\Omega^S\right)\,e_1}\right|\\
    &\geq\frac{1}{2k}\,\frac{k}{d}\sum_{n=1}^d\left|\tr{\left(\rho_{n}^S-\Omega^S\right)\,e_1}\right|\\
    &\geq\frac{1}{2d}\,\frac{1}{d_S^2}\sum_{i=1}^{d_S^2}\sum_{n=1}^d\left|\tr{\left(\rho_{n}^S-\Omega^S\right)\,e_i}\right|\\
    &\geq\frac{1}{2d\,d_S^2}\sum_{n=1}^d\sqrt{\sum_{i=1}^{d_S^2}\left[\tr{\left(\rho_{n}^S-\Omega^S\right)\,e_i}\right]^2}\\
    &\geq\frac{1}{2d\,d_S^2}\sum_{n=1}^d\sqrt{\tr{\left(\rho_n^S-\Omega^S\right)^2}}\\
    &\geq\frac{1}{2d\,d_S^2}\sum_{n=1}^d\frac{1}{\sqrt{d_S}}\tr{\sqrt{\left(\rho_n^S-\Omega^S\right)^2}}\\
    &=(d_S)^{-\frac{5}{2}}\,\frac{1}{d}\sum_{n=1}^d\,\frac{1}{2}\mathrm{tr}\left|\rho_n^S-\omega^S\right|\\
\end{split}
\end{equation}
\begin{equation}
\begin{split}  
    &=(d_S)^{-\frac{5}{2}}\frac{1}{d}\sum_{n=1}^d\D{\rho_n^S,\Omega^S}.
\end{split}
\end{equation}

If every time-averaged state $\omega$ with effective dimension at least $\frac{d}{4}$ is such that $\D{\omega^S,\Omega^S}\geq \varepsilon$, it follows that

\begin{equation}
    \frac{1}{d}\sum_{n=1}^d\D{\rho_n^S,\Omega^S} \leq (d_S)^{\frac{5}{2}}\varepsilon.
\end{equation}

\begin{flushright}
$\square$
\end{flushright}

\section{Lemma for choosing eigenstates}
In the proof of Theorem 2, we use a result to lower-bound the contribution from a single measurement outcome, $M_1$, to the distinguishability of the maximum mixture of an optimal subset of energy eigenstates in the band from the microcanonical state:
\begin{equation}
\left| \sum_{j=1}^{k}\tr{M_1\left(\rho_{n_j}-\Omega\right)}\right| \geq \frac{k}{d}\sum_{n=1}^{d}\left|\tr{M_1\left(\rho_{n}-\Omega\right)}\right|,
\end{equation}
where $k \leq \frac{d}{3}$. This is in fact a straightforward property of the finite collection $
    \left\{ \tr{M_1\left(\rho_{n}-\Omega\right)} : 1\leq n \leq d \right\}
$ of real numbers, which satisfy the property that
\begin{equation}
\sum_{n=1}^{d}\tr{M_1\left(\rho_{n}-\Omega\right)} =0.
\end{equation}
As such, we frame the result in general terms, without reference to the specific relevance of those numbers.
\vspace{1em}

\textbf{Lemma}
Let $a_1 ,..., a_d\in\mathbb{R}$ be such that $\sum_{n=1}^d a_n = 0$, and let $k\leq\frac{d}{3}$ be an integer. Then there exists a subset $\{a_{n_1},...,a_{n_k}\}$ such that $\left|\sum_{j=1}^{k}a_{n_j}\right|\geq \frac{k}{d}\sum_{n=1}^{d}\left| a_n\right|$.

\vspace{1em}
\textbf{Proof}
Without loss of generality, the $a_n$ may be ordered by value:
\begin{equation}
    a_1\geq a_2\geq ... \geq a_N\geq 0\geq a_{N+1}\geq ...\geq a_d,
\end{equation}
and, furthermore, we may assume that $N\leq\frac{d}{2}$, since swapping each $a_n\rightarrow -a_n$ does not affect sums over $\left|a_n\right|$. From this it follows that
\begin{equation}
    \sum_{n=1}^{N} a_n = -\sum_{n=N+1}^{d} a_n = \frac{1}{2}\sum_{n=1}^{d}\left| a_n\right|
\end{equation}
and
\begin{equation}
\begin{split}
    \frac{1}{N}\sum_{n=1}^{N} a_n &= \frac{1}{2N}\sum_{n=1}^{d}\left| a_n\right|\\
    &\geq \frac{1}{d}\sum_{n=1}^{d}\left| a_n\right|,
\end{split}
\end{equation}
that is, the average absolute value across $a_1,...,a_N$ is at least equal than that of $a_1,...,a_d$.

\vspace{1em}
\textbf{Case 1.} $k\leq N$
\begin{equation}
    \sum_{n=1}^k a_n \geq \frac{k}{N}\sum_{n=1}^N a_n \geq \frac{k}{d}\sum_{n=1}^d \left|a_n\right|
\end{equation}

\textbf{Case 2.} $N\leq k\leq \frac{d}{3}$
\begin{equation}
\begin{split}
    \sum_{n=1}^k a_n &= \sum_{n=1}^N a_n + \sum_{n=N+1}^k a_n\\
    &\geq \sum_{n=1}^N a_n + \frac{k-N}{d-N}\sum_{n=N+1}^d a_n\\
    &\geq \left(1-\frac{k-N}{d-N}\right)\frac{1}{2}\sum_{n=1}^d \left|a_n\right|\\
    &\geq \left(1-\frac{k}{d}\right)\frac{1}{2}\sum_{n=1}^d \left|a_n\right|\\
    &\geq \left(\frac{2k}{d}\right)\frac{1}{2}\sum_{n=1}^d \left|a_n\right|\\
    &=\frac{k}{d}\sum_{n=1}^d \left|a_n\right|\\
\end{split}
\end{equation}

So, provided $k\leq \frac{d}{3}$ and with appropriate ordering of the indices, $\left|\sum_{n=1}^{k}a_{n}\right|\geq \frac{k}{d}\sum_{n=1}^{d}\left| a_n\right|$, and the claim holds.
\begin{flushright}
$\square$
\end{flushright}
\section{Idealisations}

We now present extensions of our results for situations where some of the mathematical idealisations are relaxed. Namely, we consider states whose energy distributions have tails outside the band defining the microcanonical state, and systems with degenerate energy spectra.

\vspace{1em}
\textbf{Energy distributions with tails.}

Consider a state with time average $\omega$ satisfying that
\begin{equation}
    \tr{\Pi_\Delta\,\omega\,\Pi_\Delta} = 1-\delta,
\end{equation}
where $\Pi_\Delta$ is the projector onto an energy band of dimension d. Denote by $\mathrm{D}_{\mathcal{M}}^{\mathrm{mean}}$ and $\mathrm{D}_{\mathcal{M}}^{\mathrm{RMS}}$ the mean and root mean square distinguishability of energy eigenstates from the microcanonical state in that band, as in equations \eqref{Dmean} and \eqref{DRMS}. Furthermore, denote the time-averaged state's renormalised projection onto the energy band by
\begin{equation}
    \omega^\Delta = \frac{\Pi_\Delta\,\omega\,\Pi_\Delta}{1-\delta} 
\end{equation}
and the projection onto the band's complement by
\begin{equation}
    \omega^C = \frac{(\mathbb{I}-\Pi_\Delta)\omega(\mathbb{I}-\Pi_\Delta)}{\delta},
\end{equation}
so that the time-averaged state can be decomposed as
\begin{equation}
    \omega = \delta\,\omega^C + (1-\delta)\omega^\Delta.
\end{equation}
The distinguishability of the time-averaged state $\omega$ from the band's microcanonical state $\Omega^\Delta$ can be bounded using Theorem 1:
\begin{equation}
    \begin{split}
    \DEm{\omega,\Omega^\Delta} & = \DEm{\delta\,\omega^C + (1-\delta)\omega^\Delta,\,\Omega^\Delta}\\
    &\leq \delta\,\DEm{\omega^C,\Omega^\Delta} + (1-\delta)\,\DEm{\omega^\Delta,\Omega^\Delta}\\
    &\leq \delta + \DEm{\omega^\Delta,\Omega^\Delta}\\
    &\leq \delta + \mathrm{D}_{\mathcal{M}}^{\mathrm{RMS}}\sqrt{\frac{d}{\deff{\omega^\Delta}}-1}\\
    &\leq \delta + \sqrt{\mathrm{D}_{\mathcal{M}}^{\mathrm{mean}}\left(\frac{d}{\deff{\omega^\Delta}}-1\right)}.
    \end{split}
\end{equation}

It follows that the above bound also holds for any state $\rho$ satisfying that $\tr{\Pi_\Delta\,\rho\,\Pi_\Delta} \geq 1-\delta$.

A similar adaptation of Theorem 2 is straightforward. Let $\mathcal{S}$ be the set of time-averaged states $\omega$ satisfying the following properties:

\renewcommand{\labelenumi}{\roman{enumi}}
\begin{enumerate}
\item{\hspace{3mm}$\tr{\Pi_\Delta\,\omega\,\Pi_\Delta} \geq 1-\delta$}
\item{\hspace{3mm}$\deff{\omega^\Delta}\geq \frac{d}{4}$.}
\end{enumerate}
If every $\omega\in\mathcal{S}$ is such that $\DEm{\omega,\Omega^\Delta}\leq \varepsilon$ then the mean distinguishability of energy eigenstates in the band from the microcanonical state, $\mathrm{D}_{\mathcal{M}}^{\mathrm{mean}} \leq N_\mathcal{M}\,\varepsilon$ (or $d_S^{\frac{5}{2}}\varepsilon$, as appropriate).

These bounds follow as a direct consequence of Theorem 2, since the set $\mathcal{S}$ \textit{includes} all those states contained entirely within the energy band which have effective dimension at least $\frac{d}{4}$.

\vspace{1em}
\textbf{Degenerate spectra.}

If the Hamiltonian has degeneracies, then the situation becomes more complicated, as the energy basis is not unique (we can choose arbitrary bases inside each degenerate subspace). Consider a Hamiltonian of the form $H= \sum_m E_m \Pi_m$, where $\Pi_m$ is the projector onto the energy eigenspace with energy $E_m$. In this case, for initial mixed states, the effective dimension of a time-averaged state is given by \cite{Short_2011}
\begin{equation}
    \deff{\omega} = \frac{1}{\sum_m (\tr{\Pi_m \omega})^2 }.
\end{equation}
Given an eigenedecomposition $\omega = \sum_n p_n \rho_n$ of $\omega$, we will first show that 
\begin{equation} \label{eq:degencase2} 
\frac{1}{\sum_n p_n^2} \geq  \deff{\omega} \geq \frac{1}{g \sum_n p_n^2},
\end{equation} 
where $g$ is the degeneracy of the most degenerate energy subspace. This is because mixing inside degenerate subspaces does not change $\deff{\omega}$. To prove the first inequality, note that 
\begin{equation} \label{eqdegen1}  
\sum_n p_n^2 \leq \sum_m (\sum_{n: H \rho_n = E_m \rho_n } p_n)^2 = \frac{1}{\deff{\omega}}. 
\end{equation} 
To prove the second inequality, note that 
\begin{align}
\sum_m (\tr{\Pi_m \omega})^2 & = \sum_m (\tr{\Pi_m \Pi_m \omega})^2 \nonumber \\
& \leq \sum_m \tr{\Pi_m^2} \tr{\Pi_m \omega^2 \Pi_m} \nonumber \\
& =  \sum_m \tr{\Pi_m} \tr{\Pi_m \omega^2 } \nonumber \\
& \leq g\sum_m \tr{\Pi_m \omega^2 } \nonumber \\
& = g\,\tr{\omega^2} \nonumber \\
& = g \sum_n p_n^2
\end{align}
where in the second line we have used the Cauchy-Schwartz inequality for the Hilbert Schmidt inner product ( $(\tr{A^{\dagger} B})^2 \leq \tr{A^{\dagger} A} \tr{B^{\dagger} B}$). 

The proofs for theorems 1 and 2 are very similar to before. For theorem 1, we expand $\omega$ in its eigenbasis, and use the result of (D8) to include an additional  inequality in the final line of \eqref{eq:thm1proof}. This bounds the distinguishability of  $\omega$ from the microcanonical state, provided that the average distinguishability $D_{\mathcal{M}}^{\textrm{mean}}$ is computed in the the eigenbasis of $\omega$. 

To obtain a basis-independent result, we replace the mean distinguishability by its maximum over choices of energy basis:

\begin{equation}
    \mathrm{D}_{\mathcal{M}}^{\mathrm{mean, max}} = \max_\mathcal{B}\frac{1}{d}\sum_{n=1}^{d} \DEm{\rho_n,\Omega}.
\end{equation}

It then holds for any time-averaged state in the band that

\begin{equation}
    \DEm{\omega,\Omega} \leq \sqrt{\mathrm{D}_{\mathcal{M}}^{\mathrm{mean, max}}\left(\frac{d}{\deff\omega}-1\right)}.
\end{equation}

For theorem 2, we construct the state $\omega$, which is an equal mixture of $k \geq \frac{d}{4}$ different energy eigenstates, in the energy eigenbasis $\mathcal{B}$ which achieves the maximum in \eqref{Dmean}. The proofs then proceed as before with the only change being that the effective dimension of the state $\omega$ is no longer equal to $k$. Instead, it follows from \eqref{eq:degencase2} that 
\begin{equation}
\deff{\omega} \geq \frac{k}{g} \geq \frac{d}{4g}
\end{equation} 
We can therefore modify Theorem 2 to the following:  
\begin{theorem}
Suppose that, in an energy band with dimension $d$ and maximum energy degeneracy $g$, every time-averaged state $\omega$ with $\deff{\omega}\geq\frac{d}{4g}$ is such that $\DEm{\omega,\Omega}\leq\varepsilon$ for some $\varepsilon>0$.

\renewcommand{\labelenumi}{\roman{enumi}}
\begin{enumerate}
    \item If the number of outcomes across all available measurements, $N_{\mathcal{M}}=\sum_{M\in\mathcal{M}}N(M)$ is finite, then the mean distinguishability of energy eigenstates from the band's microcanonical state,
\begin{equation}
 \mathrm{D}_{\mathcal{M}}^{\mathrm{mean, max}}\leq N_{\mathcal{M}}\,\varepsilon.
\end{equation}
    \item{Furthermore, if $\mathcal{M}$ represents the set of all POVMs on a subsystem of dimension $d_\textrm{S}$, the mean distinguishability is bounded by
\begin{equation}
    \mathrm{D}_{\mathcal{M}}^{\mathrm{mean, max}}\leq {d_\textrm{S}}^\frac{5}{2}\varepsilon.
\end{equation}}
\end{enumerate}
\end{theorem}
Note that some change to Theorem 2 is inevitable in the presence of degeneracies, because if all energy subspaces have degeneracy $g$, then the maximum value that $\deff{\omega}$ could take is $\frac{d}{g}$. If $g>4$ there would be no states satisfying $\deff{\omega} \geq \frac{d}{4}$.

\end{document}